\documentstyle[aps,12pt]{revtex}
\newcommand{\bb}{\begin{equation}}
\newcommand{\ee}{\end{equation}}
\newcommand{\ba}{\begin{array}}
\newcommand{\ea}{\end{array}}

\begin {document}
\baselineskip 2.2pc

\title{On the partial wave amplitude of Coulomb scattering
       in three dimensions\thanks{published in Am. J. Phys.
       {\bf 68} (2000) 1056-1057.}}
\author{Qiong-gui Lin\thanks{E-mail:
        qg\_lin@163.net, stdp@zsu.edu.cn}}
\address{Department of Physics, Zhongshan University, Guangzhou
        510275,\\
        People's  Republic of China\thanks{Mailing address}\\
        and\\
        China Center of Advanced Science and Technology (World
	Laboratory),\\
        P.O.Box 8730, Beijing 100080, People's Republic of China}

\maketitle
\vfill

\begin{abstract}
{\normalsize The partial wave series for the Coulomb scattering
amplitude in three dimensions is evaluated in a very simple
way to give the  closed result.}
\end{abstract}
\vfill
\leftline {PACS number(s): 03.65.Nk}
\newpage

The scattering of particles by a Coulomb field is an important
subject in quantum mechanics as well as in classical mechanics,
because the Coulomb field describes real physical interations
between charged particles, say, electrons and nucleuses. In the
nonrelativistic quantum mechanics, the problem is described by a
Schr\"odinger equation. In ordinary three-dimensional space, this
equation can be solved in rotational parabolic coordinates, and the
scattering amplitude is obtained in a closed form.
As a special case of a centrally symmetric field, the Schr\"odinger
equation can also be solved in the spherical coordinates, and the
scattering amplitude is obtained as a partial wave series. 
This has been well known for a long time, and the solutions can be
found in the popular textbooks of quantum mechanics [1,2].
However, it appears that the partial wave amplitude is not
evaluated in the textbooks. Were it not for the solution in the
parabolic coordinates, how can one directly sum up the partial wave
series to obtain the closed form? A solution can be found in an early
paper of Mott [3]. However, the calculation is not well known
since it is not given even in the famous monograph on collision theory
by that author and Massey [4]. Other books on scattering theory 
do not deal with the matter either [5,6]. The purpose of this note
is to evaluate the partial wave series in a simpler way. The method
is easy for the students and thus may be of some interest for
the pedagogical purpose.

Let us write the Hamiltonian in the form
\bb
H={{\bf p}^2\over 2\mu}-{\kappa\over r},
\ee          
where $\mu$ is the reduced mass of the system, 
and $\kappa>0$ ($\kappa<0$) corresponds to an attractive (repulsive)
force. The partial wave scattering amplitude obtained by solving the
problem in spherical coordinates reads [1,2]
\bb
f(\theta)={1\over 2ik}\sum_{l=0}^\infty (2l+1)(S_l-1)P_l(\cos\theta),
\ee     
where $k=\sqrt{2\mu E}/\hbar$ ($E>0$ is the energy of the incident
particle), and
\bb
S_l=\exp(2i\delta_l)={\Gamma(l+1-i\beta)\over\Gamma(l+1+i\beta)},
\ee     
where $\delta_l$ is the phase shift of the $l$th partial wave, and
$\beta=\mu\kappa/\hbar^2 k=\kappa/\hbar v$ where $v$ is the incident
velocity. Since [1,2]
\bb
\sum_{l=0}^\infty (2l+1)P_l(\cos\theta)=4\delta(\cos\theta-1),
\ee     
we have for $\theta\ne 0$
\bb
f(\theta)={1\over 2ik}\sum_{l=0}^\infty (2l+1)S_l P_l(\cos\theta).
\ee     
We define
\bb
g(x)=\sum_{l=0}^\infty (2l+1)S_l P_l(x),
\ee     
and are going to evaluate it. Using the relation
\bb
(2l+1)P_l(x)=P'_{l+1}(x)-P'_{l-1}(x),
\ee     
where $P_{-1}(x)\equiv 0$, and define an auxiliary function
\bb
G(x)=\sum_{l=0}^\infty S_l [P_{l+1}(x)-P_{l-1}(x)],
\ee     
we have
\bb
g(x)=G'(x).
\ee     
Next using the ralation
\bb
(2l+1)xP_l(x)=(l+1)P_{l+1}(x)+lP_{l-1}(x),
\ee     
and the definition (8), we have
\bb
xg(x)=\sum_{l=0}^\infty S_l [(l+1-i\beta)P_{l+1}(x)+
(l+i\beta)P_{l-1}(x)]+i\beta G(x).
\ee     
Then making use of the relation
\bb
\Gamma(z+1)=z\Gamma(z),
\ee     
and Eq. (3), we have
\bb
(l+1-i\beta)S_l=(l+1+i\beta)S_{l+1},\quad
(l+i\beta)S_l=(l-i\beta)S_{l-1}.
\ee     
Substituting these relations into Eq. (11), after some simple
algebras we have
\bb
xg(x)=g(x)+i\beta G(x)-i\beta S_0.
\ee     
Taking Eq. (9) into account, we obtain
\bb
(1-x)G'(x)+i\beta G(x)=i\beta S_0.
\ee     
This is an ordinary differential equation for the auxiliary function
$G(x)$. It can be easily solved to give
\bb
G(x)=A\exp[i\beta\ln (1-x)]+S_0,
\ee     
where $A$ is a constant of integration. To determine $A$ we consider
the series (8) at $x=-1$. Note that $P_l(-1)=(-1)^l$ and
$P_{-1}(x)\equiv 0$, only the term with $l=0$ contributes and
so $G(-1)=-S_0$. Substituting this into the above result one can find
$A$ and arrives at
\bb
G(x)=-2S_0\exp[i\beta\ln ((1-x)/2)]+S_0.
\ee     
It should be remarked that in getting Eq. (9) we have differentiated
$G(x)$ term by term. This is not allowed at $x=1$ since 
$g(x)$ is obviously singular at the  point. Thus Eq. (15) does not
hold at $x=1$, and it is not surprising to find that the result
(17) does not agree with the definition (8) at that point. By using
Eq. (9) we can easily find $g(x)$. Setting $x=\cos\theta$ and
multiplying it by the coefficient $1/2ik$ we arrive at
\bb
f(\theta)={\Gamma(1-i\beta)\over i\Gamma(i\beta)}{\exp[i\beta\ln
\sin^2(\theta/2)]\over 2k\sin^2(\theta/2)}.
\ee     
This is the closed result obtained in the parabolic coordinates [1,2].
Two remarks about the result. First, we have dropped a
$\delta(\cos\theta-1)$ term for $\theta\ne 0$ in the above
calculation. Here the final result is more singular than
$\delta(\cos\theta-1)$ when $\theta\to 0$, thus that term can indeed
be dropped everywhere. Second, it has been argued that $f(\theta)$
in the form of Eq. (2) or (5) is divergent even when $\theta\ne 0$ [7].
The final result (18) is well defined everywhere except at $\theta=0$,
however. Therefore the above calculation may also be regarded as a
regularization procedure. (The main point lies in the term by term
differentiation pointed out above, which is mathematically not
allowed even when $x\ne 1$.) The validity of this procedure is obvious
because the final result (18) coincides with the one obtained from
exact solutions in the parabolic coordinates. It should be pointed out
that the calculation of Mott [3], though rather different from ours,
involves some similar story.

Mathematically one can consider the Coulomb scattering problem 
in a general space dimension $d\ge 2$. In $d=2$ the problem has been
solved in both parabolic coordinates [8-9] and polar coordinates [10].
The scattering amplitude is obtained in a closed form in the former
case and as a partial wave series in the latter case. The method for
summing up the partial wave series in $d=2$ is rather different from
that in $d=3$ [10,11]. In more general dimensions $d>3$ the problem
has been solved in parabolic coordinates and a closed form for the
scattering amplitude was obtained [12]. If one solves the problem
in $d$-dimensional spherical coordinates, one would obtain for the
scattering amplitude a partial wave series similar to Eq. (2) , where
the Legendre polynomials are replaced by the Gegenbauer ones. We
expect that the method given by Mott or the one given here may be
efficient for summing up that series as well.

\vskip 1pc

The author is grateful to Professor W. Dittrich and Professor B. R.
Holstein for communications and encouragement.
This work was supported by the
National Natural Science Foundation of China.

\end{document}